\documentclass[conference]{IEEEims}
\input epsf
\usepackage{graphicx}

\begin{document}

\title{\LARGE Experimental Realization of an NMR Quantum
Switch}

\author{\authorblockN{
I-Ming Tsai\authorrefmark{1}, Sy-Yen Kuo\authorrefmark{1},
Shou-Lin Huang\authorrefmark{2}, Ying-Chih Lin\authorrefmark{2},
Tso-Tsai Chen\authorrefmark{1}}
\vspace{0.1 in}
\authorblockA{\authorrefmark{1}Department of Electrical
Engineering\\
National Taiwan University, Taipei, Taiwan\\
E-mail: sykuo@cc.ee.ntu.edu.tw}
\vspace{0.05 in}
\authorblockA{\authorrefmark{2}Department of Chemistry\\
National Taiwan University, Taipei, Taiwan\\
E-mail: yclin@ntu.edu.tw}}

\maketitle

\begin{abstract}
In this paper, we report an experimental realization of quantum
switch using nuclear spins and magnetic resonant pulses. The
nuclear spins of $^{1}$H and $^{13}$C in carbon-13 labelled
chloroform are used to carry the information, then nuclear
magnetic resonance pulses are applied to perform either bypass or
cross function to achieve the switching. Compared with a
traditional space or time domain switch, this switching
architecture is much more scalable, therefore a high throughput
switching device can be built simply by increasing the number of
I/O ports. In addition, it can be used not only as a device to
switch classical information, but also a building block of quantum
information networks.
\end{abstract}

\IEEEoverridecommandlockouts


\IEEEpeerreviewmaketitle

\section{Introduction}

The study of quantum information science has expanded rapidly due
to the discovery of several efficient quantum algorithms
\cite{Deu92}\cite{Sho94}\cite{Gro96} and the progress in physical
implementation schemes. Recent advances in quantum mechanical
technology like ion traps \cite{Cir95}, optical cavities
\cite{Tur95}, nuclear magnetic resonance \cite{Ger97}, quantum
dots \cite{Los98}, and silicon-based solutions \cite{Kan98}, have
brought scientists a number of ways to realize quantum mechanical
applications. In these solutions, Nuclear Magnetic Resonance (NMR)
is a mature technology which has wide spread applications in
chemistry and medical areas. There have been many reports that
demonstrate NMR is capable of implementing a quantum computation
system and solving intractable problems
\cite{Chu98a}\cite{Jon98}\cite{Chu98}\cite{Van98}. In this paper,
we report an experimental realization of 2$\times$2 NMR quantum
switch using $^{1}$H and $^{13}$C in carbon-13 labelled chloroform
($^{13}$CHCl3) as the information carrying units.

Digital switching, in general, can be categorized into two major
classes - circuit switching and packet switching. For circuit
switching, the switching module moves the data in each time slot
between the input and output ports. As for packet switching, a
packet with either a fixed or variable length is forwarded to the
destination according to its header. There are many underlying
architectures that can be used to provide the function of
switching. For example, time domain switching and space domain
switching are two commonly used architectures. In our previous
work \cite{Tsa02}, we have presented a new switching architecture
such that digital data can be switched in the quantum domain. The
proposed mechanism supports unicasting as well as multicasting,
and is scalable and non-blocking. It can be used to build
classical and quantum information networks.

Note that, as long as the state of the information carrier can be
manipulated according to quantum mechanics, the proposed switching
architecture is independent of the underlying technology. In this
paper we demonstrate a 2$\times$2 quantum switch using nuclear
spins and NMR pulses. More specifically, the nuclear spins of
$^{1}$H and $^{13}$C in carbon-13 labelled chloroform are used to
carry the information to be processed, then RF pulses are applied
to perform either bypass or cross action to achieve the function
of switching. Compared with a traditional space or time domain
switch, this switching architecture is much more scalable. For an
n$\times$n quantum switch, the space consumption is linear and the
time complexity is constant. With these advantages, a high
throughput switching device can be built simply by increasing the
number of I/O ports.

\section{Circuit Implementation of a Quantum Switch}

\subsection{Quantum Circuits}

\begin{figure}[htbp]
 \center
 \scalebox{0.27}{\includegraphics{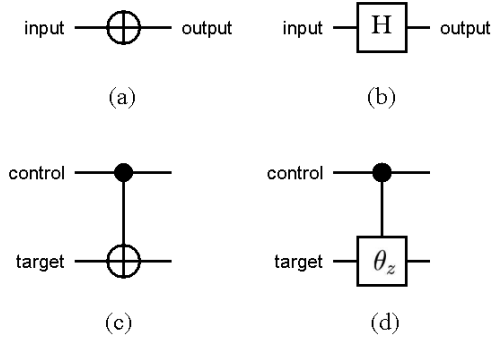}}
 \caption{The symbol for a (a) Not (b) Hadamard (c) Control-not (d) Control-phase-shift gate}
 \label{not_and_cn}
\end{figure}

For the quantum state of a nuclear spin, there are two eigen
states, denoted by $\vert 0 \rangle$ for spin-up and $\vert 1
\rangle$ for spin-down. The state of each spin can be written as a
linear combination of these two eigen states, so we have the state
$\vert \psi \rangle$ of a qubit as
\begin{equation}
\vert \psi \rangle = c_0 \vert 0 \rangle + c_1 \vert 1 \rangle \ ,
\end{equation}
with $c_0$, $c_1 \in {\cal C}$ and $\vert c_0 \vert^2 + \vert c_1
\vert^2 = 1$. The state described above exhibits an unique
phenomenon in quantum mechanics called {\em superposition}. When a
particle is in such a superposed state, it has a part
corresponding to $\vert 0 \rangle$ and a part corresponding to
$\vert 1 \rangle$, at the same time. To distinguish the above
system from the classical binary logic, a unit that carrying
binary information in such a quantum system is referred to as a
quantum binary digit, or {\em qubit}. A quantum system can be
manipulated in many different ways. These operations are generally
called {\em quantum gates}. A quantum gate must be unitary in its
matrix form and can be pictorially described as a rotation on a
sphere. For example, a quantum {\bf Not} ({\bf \small N}) gate
applied to a single qubit is actually a rotation of $\pi$ around
the ${\hat x}$ axis. The symbol of an {\bf \small N} gate is shown
in Fig.\ref{not_and_cn}(a). Note that the horizontal line
connecting the input and the output is not a physical wire,
instead it represents a qubit under time evolution. Another
important gate is the {\bf Hadamard} ({\bf \small H}) gate, as
shown in Fig.\ref{not_and_cn}(b). The {\bf \small H} gate performs
a rotation of $\pi$ around the axis of ($x=z; \ y=0$) and is
actually an exchange of ${\hat x}$ axis with ${\hat z}$ axis.

Two or more qubits can also form a quantum system jointly. A
two-qubit system is spanned by
$\vert 00 \rangle$, $\vert 01 \rangle$,
$\vert 10 \rangle$, and $\vert 11 \rangle$, i.e.
\begin{equation}
\vert \psi \rangle = c_0 \vert 00 \rangle + c_1 \vert 01
\rangle + c_2 \vert 10 \rangle + c_3 \vert 11 \rangle \ ,
\end{equation}
with $c_0$, $c_1$, $c_2$, $c_3 \in {\cal C}$ and
$\vert c_0 \vert^2 + \vert c_1 \vert^2 + \vert c_2 \vert^0 + \vert
c_3 \vert^2 = 1$. In general, the space of a multi-qubit system is spanned by the
basis of the tensor product of each spaces. An example of two-qubit gate
is the {\bf Control-Not} ({\bf \small CN}) gate, as shown in
Fig.\ref{not_and_cn}(c). A {\bf \small CN} gate consists
of one {\em control} bit $x$ and one {\em target} bit $y$.
The target qubit will be inverted only when the control qubit is $\vert 1 \rangle$.
Assuming $x$ is the control bit, the gate can be written
as {\bf \small CN}($\vert x,y \rangle$)$=\vert x,x \oplus y \rangle$, where
$\oplus$ denotes exclusive-or. A generalization of the {\bf \small CN}
gate is the {\bf Control-phase-shift} gate, which shifts the phase of the target
qubit for $\theta$ only when the control qubit is $\vert 1 \rangle$.
A control-phase-shift gate is shown in Fig.\ref{not_and_cn}(d).

Although we can define and implement a multi-qubit gate in a multi-qubit
system, it has been shown
that two-qubit gates are sufficient to implement any unitary
operation \cite{Div95}\cite{Bar95.1}\cite{Bar95.2}.
In practice, a given multi-qubit quantum operation can be implemented
using only {\bf \small CN} gates and single qubit rotations.

\subsection{NMR Implementation}

Among various physical implementation schemes, NMR is one of the
technologies to provide us with a way to achieve the desired
quantum operations. When a nucleus with a non-zero spin is placed
in a magnetic field (the ${\hat z}$ axis, by convention), the spin
aligns in either the same direction (${\hat z}$ axis) or in the
opposite direction ($-{\hat z}$ axis). A nucleus with its spin
aligned with the field has lower energy than the nucleus with its
spin aligned in the opposite direction. In addition to the
alignment, the spin {\em precesses} with a {\em Larmor frequency}
of
\begin{equation}
\omega_{0} = - \gamma B_{0} \ ,
\end{equation}
where $B_{0}$ is the magnetic field and $\gamma$ is called {\em
magnetogyric ratio} of that nucleus. In such a system, an nuclear
spin can be used as a qubit, and a quantum operation can be
implemented using a rotating magnetic field which is generated by
NMR Radio Frequency (RF) pulses.

More specifically, a single qubit operation is accomplished by
applying a magnetic field ($B_{1}$, by convention) in the rotating
frame. The quantum state is manipulated by this RF signal and the
net magnetization is tipped away from the ${\hat z}$ axis. Since
the tipping angle is proportional to the power and duration of the
pulse, the state can be observed via the direction of the net
magnetization. In practice, the power is usually fixed and the
duration of the pulse is changed to control the state evolution.
In other words, the duration for a $\pi$ pulse is twice as that of
a $\frac{\pi}{2}$ pulse and any other rotations can be calculated
accordingly. An example of manipulating the state of a single
qubit is to use RF pulses along the ${\hat x}$ axis in the
rotating frame. Such a pulse can be written as
\begin{equation}
e^{-i\theta I_{x}} \ ,
\end{equation}
where $I_{x}$ is defined as $\frac{\sigma_{x}}{2}$. When $\theta =
\pi$, it is a quantum {\bf \small N} gate, up to an unimportant
global phase difference. This means a $\pi$ pulse along the ${\hat
x}$ axis actually rotates the state of a single qubit by $\pi$
around the ${\hat x}$ axis. Similarly, an RF pulse along the
${\hat y}$ axis or the ${\hat z}$ axis can be written as
$e^{-i\theta I_{y}}$ and $e^{-i\theta I_{z}}$, with $I_{y}$ and
$I_{z}$ defined as $\frac{\sigma_{y}}{2}$ and
$\frac{\sigma_{z}}{2}$ respectively. In addition, pulses along the
$-{\hat x}$, $-{\hat y}$, or $-{\hat z}$ axes can be implemented
in the same way with an effect of negative $\theta$. This can be
written as, for example,
\begin{equation}
e^{-i\theta I_{-x}} = e^{-i(-\theta)I_{x}} .
\end{equation}
Note that a rotation around the ${\hat z}$ axis (i.e. $e^{-i\theta
I_{z}}$) is actually a phase-shift and can be implemented in a
variety of ways. For example, since
\begin{equation}
e^{-i\theta I_{z}} = e^{-i \frac{\pi}{2} I_{-x}} \cdot e^{-i
\theta I_{y}} \cdot e^{-i \frac{\pi}{2} I_{x}} \ ,
\end{equation}
a ${\hat z}$ pulse can be decomposed into three pulses along the
${\hat x}$ and ${\hat y}$ axes as follows:
\begin{equation}
(\theta)_{I_{z}} \Leftrightarrow (\frac{\pi}{2})_{I_{-x}} -
(\theta)_{I_{y}} - (\frac{\pi}{2})_{I_{x}} \ . \label{z-pulse}
\end{equation}
Similarly, an {\bf \small H} gate can be implemented as
\begin{eqnarray}
H & = & e^{-i \frac{\pi}{4} I_{y}} \cdot e^{-i \pi I_{x}} \cdot e^{-i \frac{\pi}{4} I_{-y}}\\
& \Rightarrow & (\frac{\pi}{4})_{I_{y}} - (\pi)_{I_{x}} -
(\frac{\pi}{4})_{I_{-y}} \ .
\end{eqnarray}

\begin{figure}[htbp]
 \center
 \scalebox{0.2}{\includegraphics{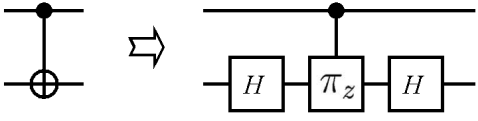}}
 \caption{The decomposition of a control-not gate}
 \label{cn-decompose}
\end{figure}

Two-qubit conditional logic is achieved by the effect of spin-spin
coupling. In this experiment, the spin-spin coupling is the
interaction between the spin states of $^{1}$H and $^{13}$C,
measured in Hertz (Hz). Due to this coupling, the precession of
one qubit can be increased or decreased, depending on the state of
the other qubit. If a specific duration of free evolution is
designed into the pulse sequence, the state of the target qubit
can be phase-shifted conditionally on the state of the control
qubit. This is essentially the control-phase-shift gate described
before and can be used to generate other quantum conditional logic
gates. An example of two-qubit conditional logic is the {\bf
\small CN} gate, which can be decomposed into two {\bf \small H}
gates and a control-phase-shift of $\pi$, as shown in
Fig.\ref{cn-decompose}.

Assuming the first qubit is the control qubit and the second qubit
is the target qubit, the control-phase-shift of $\pi$ in
Fig.\ref{cn-decompose} can be written as
\begin{equation}
e^{-i\frac{\pi}{2}(-\frac{1}{2}E+I_{z}+S{z}-2I{z}S{z})}\ .
\end{equation}
Each of the four terms is explained as follows. Obviously, the
first term $e^{-i\frac{\pi}{2}(-\frac{1}{2}E)}$ is an identity, so
no pulse is needed. The second term $e^{-i\frac{\pi}{2}I_{z}}$ and
the third term $e^{-i\frac{\pi}{2}S_{z}}$ are rotations around the
$ {\hat z}$ axis for the control and target qubit respectively.
These can be implemented using ${\hat x}$ and ${\hat y}$ pulses as
described in Eq.(\ref{z-pulse}). The fourth term
$e^{-i\frac{\pi}{2}(-2I{z}S{z})}$ corresponds to the free
evolution under spin-spin coupling between the control and target
qubit. This can be implemented as a delay of $\tau=\frac{1}{2J}$,
where $J$ is the spin-spin coupling constant. In summary, the NMR
pulse sequence for a {\bf \small CN} gate is as follows:
\begin{eqnarray}
& &(\pi/4)_{I_{y}}-(\pi)_{I_{x}}-(\pi/4)_{I_{-y}} \nonumber \\
& - & (\pi/2)_{S_{y}} - (\pi/2)_{S_{x}} - (\pi/2)_{S_{-y}} - \tau \nonumber \\
& - & (\pi/2)_{I_{y}} - (\pi/2)_{I_{x}} - (\pi/2)_{I_{-y}} \nonumber \\
& - & (\pi/4)_{I_{y}} - (\pi)_{I_{x}}   - (\pi/4)_{I_{-y}}
\label{cn_ps}
\end{eqnarray}

\subsection{Quantum Switch}

\begin{figure}[htbp]
 \center
 \scalebox{0.21}{\includegraphics{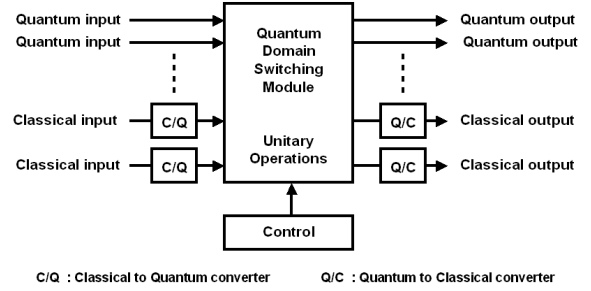}}
 \caption{The architecture of a digital quantum switch}
 \label{framework}
\end{figure}

In this experiment, the architecture of a {\em quantum switch}
\cite{Tsa02} is implemented, as depicted in Fig.\ref{framework}.
The I/O ports of a quantum switch can be configured to carry
either quantum or classical information. For those I/O ports that
carry classical information, the incoming bit stream has to be
converted into qubits for further processing. This can be done by
a classical to quantum converter (C/Q) which converts '0' into
$\vert 0 \rangle$ and '1' into $\vert 1 \rangle$. All qubits are
then permuted using unitary operations. After the permutation, all
qubits are converted back into their classical form by a quantum
to classical converter (Q/C).

\begin{figure}[htbp]
 \center
 \scalebox{0.2}{\includegraphics{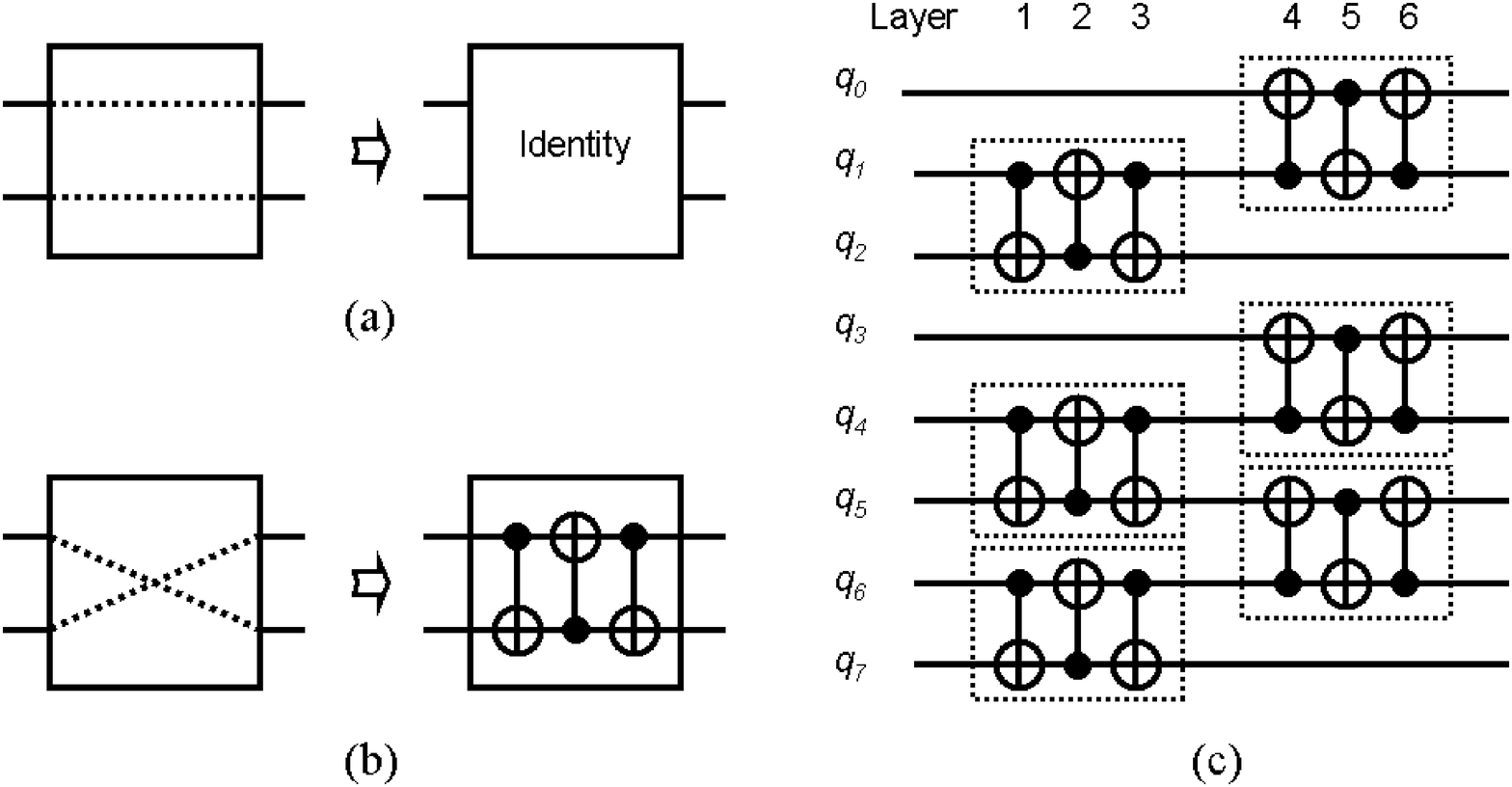}}
 \caption{A 2$\times$2 quantum switch in (a) bypass mode (b) cross mode; (c) A general 8$\times$8 quantum switch}
 \label{switch_mode}
\end{figure}

Following this principle, a 2$\times$2 quantum switch is shown in
Fig.\ref{switch_mode}(a) and Fig.\ref{switch_mode}(b). If the
quantum switch is in the bypass mode, no operation is needed.
However, for a 2$\times$2 quantum switch in the cross mode, three
{\bf \small CN} gates (with the middle one up side down) are used
to switch the quantum state. Note that an n$\times$n quantum
switch is {\em not} constructed by concatenating 2$\times$2 switch
elements as in classical Clos networks. Instead, an
n$\times$n quantum switch is implemented using $6$ (constant)
layers of {\bf \small CN} gates, which is an important feature of
this architecture. An example of an $8 \times 8$ quantum switch is
shown in Fig.\ref{switch_mode}(c).

\section{Experimental Realization}

With a Bruker Avance DMX-500MHz NMR system, carbon-13 labelled
chloroform ($^{13}$CHCl3) in d6-Acetone at room temperature is
used as our switching fabric. The $^{1}$H and $^{13}$C atoms in
chloroform are used as the information carrier. In this
experiment, channel 1 is setup for $^{1}$H and channel 2 is setup
for $^{13}$C. The observed channel is the proton spectrum in
channel 1. The C/Q is done by RF pulses to convert the classical
information into quantum states. The pulse sequences for a quantum
switch in the cross mode are formed by three {\bf \small CN} gates
as in Eq.(\ref{cn_ps}), with the second {\bf \small CN} gate up
side down.

In this experiment, the power of the pulses are set at 3.00 dB for
channel 1 ($^{1}$H) and -3.00 dB for channel 2 ($^{13}$C). The
duration for a $\frac{\pi}{2}$ pulse is 9.5 $\mu$s for channel 1
and 12.6 $\mu$s for channel 2. Since the rotation of a single qubit
quantum operation is proportional to the product of the power and
duration of the RF pulses, a simple calculation will give the
duration for a $\frac{\pi}{4}$ pulse and $\pi$ pulse. The free
evolution (under the control of spin-spin coupling) is the source
of conditional logic in quantum computing. The spin-spin coupling
between $^{1}$H and $^{13}$C is measured to be 215 Hz. As a
result, the duration of a free evolution of $\frac{\pi}{2}$ is
calculated as 1.165 $\mu$s, and a $\pi$ rotation is 2.33 $\mu$s,
twice the number for a $\frac{\pi}{2}$ rotation. These parameters
are assigned to the D, P, and PL arrays in Bruker's pulse program.

In an NMR experiment, usually a number of identical scans are
carried out to improve the signal to noise ratio (at the expense
of taking more time to carry out the experiment). This parameter
is called NS (Number of Scans). A typical NS value for a standard
chemical experiment is on the order of $1000$ to $10000$,
depending on the density of the sample. Also, to allow the sample
to reach a steady state, several dummy pulses are applied before
the real pulses, without collecting the NMR signal. This parameter
is called DS (Dummy Scans). A typical DS value is around $4$ to
$8$. Since an enriched sample is used in this experiment, the NS
and DS are set up as $8$ and $0$ respectively.

\begin{figure}[htbp]
 \center
 \scalebox{0.24}{\includegraphics{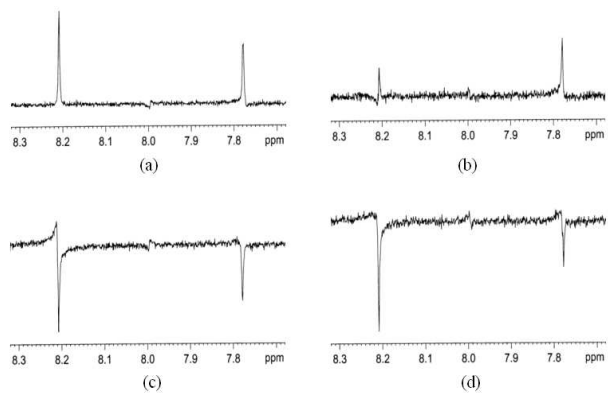}}
 \caption{The result on channel 1 with input (a) 00 (b) 10 (c) 01 (d) 11}
 \label{cross-result}
\end{figure}

Other acquisition parameters include the Time Domain size (TD)
which specifies the total number of points to be collected in the
time domain, and the FID Resolution (FIDRES) which indicates the
frequency range between each point. In this experiment, the TD is
32768 and the FIDRES is 0.305176 Hz. As a result, with an observed
Spectral Width (SW) of 10000 Hz, the acquisition time is 1.63845
sec. After the data acquisition and post-acquisition processing
(especially phase correction), the result for a quantum switch in
cross mode is shown in Fig.\ref{cross-result}.

\section{Conclusion}

In this paper, we demonstrate an experimental realization of a
quantum switch using nuclear spins and magnetic resonant pulses.
The implementation of a quantum switch once again shows us a
significant possibility of applying microscopic physics to
classical engineering problems. In our experiment, the NMR
technology provides with us a versatile control mechanism which
can be used to improve the performance of digital switching.
Although noise, loss of coherence and manufacturing issues are
still bothering scientists, judging from the technological
evolution in the past few decades, we believe there will be
tremendous technical progress in the near future.



\begin{thebibliography}{99}

\bibitem{Deu92}
D. Deutsch and R. Jozsa, "Rapid solution of problems by quantum computation,"
{\em Proc. Roy. Soc. Lond. A}, vol.439, pp. 553-558, 1992.

\bibitem{Sho94}
P. Shor, "Algorithms for quantum computation: discrete logarithms
and factoring," in {\em Proc. of the 35th Annual IEEE Symposium on
the Foundations of Computer Science}, 1994, pp. 124-134.

\bibitem{Gro96}
L. Grover, "A fast quantum mechanical algorithm for database
search," in {\em Proc. of the 28th Annual ACM Symposium on the
Theory of Computing}, 1996, pp. 212-219.

\bibitem{Cir95}
J. Cirac and P. Zoller, "Quantum computation with cold trapped
ions," {\em Phys. Rev. Lett.}, vol. 74, pp. 4091-4094, 1995.

\bibitem{Tur95}
Q. Turchette, C. Hood, W. Lange, H. Mabuchi and H. Kimble,
"Measurement of conditional phase shifts for quantum logic," {\em
Phys. Rev. Lett.}, vol. 75, pp. 4710-4713, 1995.

\bibitem{Ger97}
N. Gershenfeld and I. Chuang, "Bulk spin resonance quantum
computation," {\em Science}, vol. 275, pp. 350-356, 1997.

\bibitem{Los98}
D. Loss and D. DiVincenzo, "Quantum computation with quantum
dots," {\em Phys. Rev. A}, vol. 57, pp. 120-126, 1998.

\bibitem{Kan98}
B. Kane, "A silicon-based nuclear spin quantum computer," {\em
Nature}, vol. 393, pp. 133-137, 1998.

\bibitem{Chu98a}
I. Chuang, L. Vandersypen, X. Zhou, D. Leung and S. Lloyd,
"Experimental realization of a quantum algorithm," {\em Nature},
vol. 393, pp. 143-146, 1998.

\bibitem{Jon98}
J. Jones, M. Mosca and R. Hansen, "Implementation of a quantum
search algorithm on a quantum computer," {\em Nature}, vol. 393,
pp. 344-346, 1998.

\bibitem{Chu98}
I. Chuang, N. Gershenfeld and M. Kubinec, "Experimental
Implementation of Fast Quantum Searching," {\em Phys. Rev. Lett.},
vol. 80, no. 15, pp. 3408-3411, 1998.

\bibitem{Van98}
L. Vandersypen, M. Steffen, G. Breyta, C. Yannoni, M. Sherwood and
I. Chuang, "Experimental realization of Shor's quantum factoring
algorithm using nuclear magnetic resonance," {\em Nature}, vol.
414, pp. 883-887, 2001.

\bibitem{Tsa02}
I-Ming Tsai and Sy-Yen Kuo. "Digital Switching in the Quantum
Domain," {\em IEEE Trans. Nano.}, vol. 1 no. 3, pp. 154-164, 2002.

\bibitem{Div95}
D. DiVincenzo, "Two-bit gates are universal for quantum
computation," {\em Phys. Rev. A}, vol. 51(2), pp. 1015-1022, 1995.

\bibitem{Bar95.1}
A. Barenco, "A universal two-bit gate for quantum computation,"
{\em Proc. Roy. Soc. Lond. A}, vol. 449, pp. 679-683, 1995.

\bibitem{Bar95.2} A. Barenco, C. Bennett, R. Cleve, D. DiVincenzo, N. Margolus,
P. Shor, T. Sleator, J. Smolin, and H. Weinfurter, "Elementary
gates for quantum computation," {\em Phys. Rev. A}, vol. 52(5),
pp. 3457-3467, 1995.


\end{thebibliography}
\end{document}